\documentclass[aps,prl,twocolumn,superscriptaddress,floatfix,showpacs,groupedaddress]{revtex4-1}
\usepackage{hyperref}
\usepackage{graphicx}
\usepackage{amsfonts}
\usepackage{amssymb}
\usepackage{amsmath}
\usepackage{mathrsfs}

\usepackage{txfonts}
\usepackage{lipsum}
\usepackage{color}
\usepackage{wasysym}

\usepackage{letltxmacro}
\DeclareMathOperator{\Tr}{Tr}
\DeclareMathOperator{\Det}{Det}

\DeclareMathOperator{\Imag}{Im}
\DeclareMathOperator{\Arg}{Arg}

\begin{document}
\title{Geometrical phase shift in Friedel oscillations}
\author{C. Dutreix}
\author{P. Delplace}
\affiliation{Univ Lyon, Ens de Lyon, Univ Claude Bernard, CNRS, Laboratoire de Physique, F-69342 Lyon, France}


\begin{abstract}
This work addresses the problem of elastic scattering through a localized impurity in a one-dimensional crystal with sublattice freedom degrees. The impurity yields long-range interferences in the local density of states known as Friedel oscillations. Here, we show that the internal degrees of freedom of Bloch waves are responsible for a geometrical phase shift in Friedel oscillations. The Fourier transform of the energy-resolved interference pattern reveals a topological property of this phase shift, which is intrinsically related to the Bloch band structure topology in the absence of impurity. Therefore, Friedel oscillations in the local density of states can be regarded as a probe of wave topological properties in a broad class of classical and quantum systems, such as acoustic and photonic crystals, ultracold atomic gases in optical lattices, and electronic compounds.
\end{abstract}

\maketitle

\begin{figure}[t]
\centering
$\begin{array}{cc}
~~\includegraphics[trim = 00mm 97.1mm 0mm 100mm, clip, width=3.5cm]{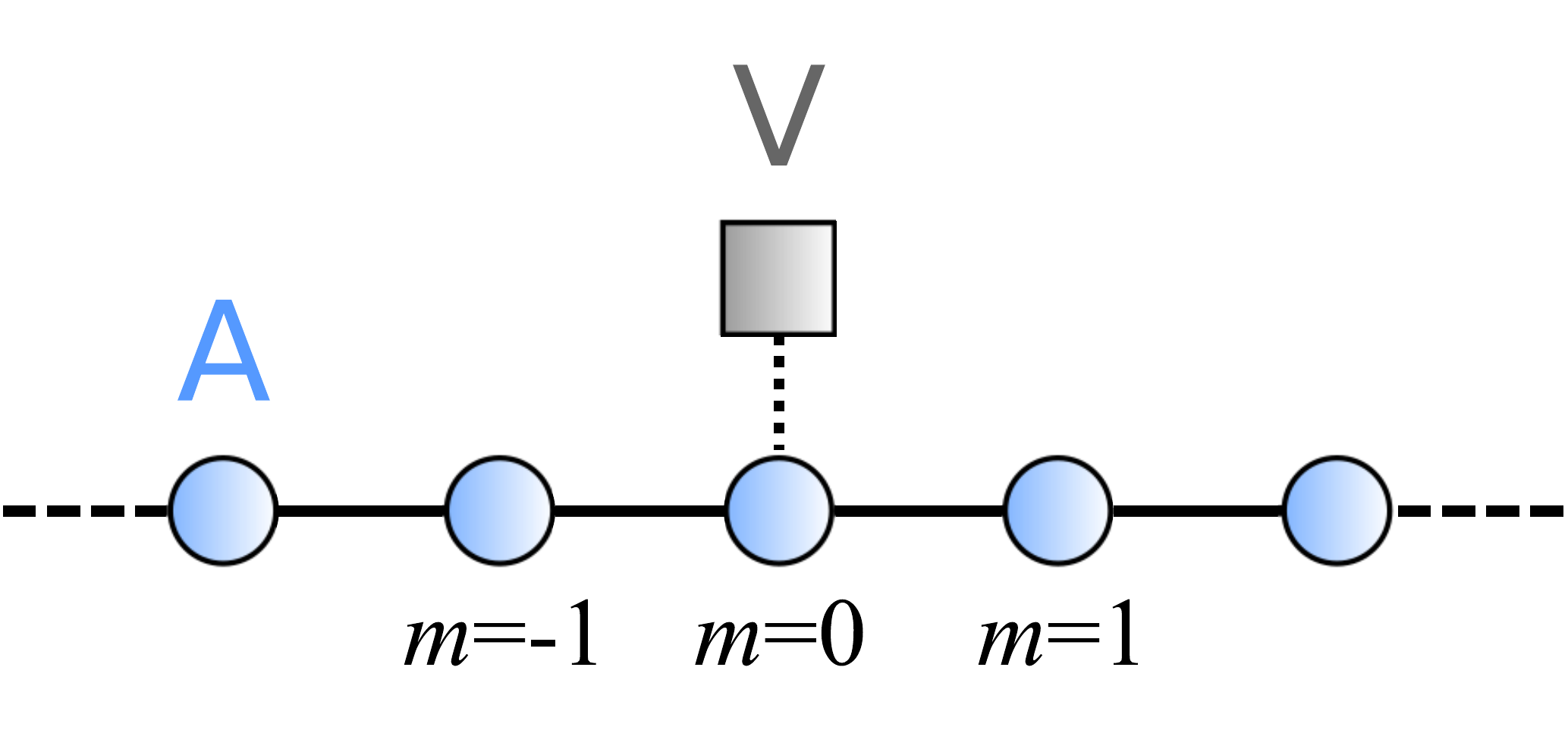}&
~~\includegraphics[trim = 00mm 100mm 0mm 100mm, clip, width=3.5cm]{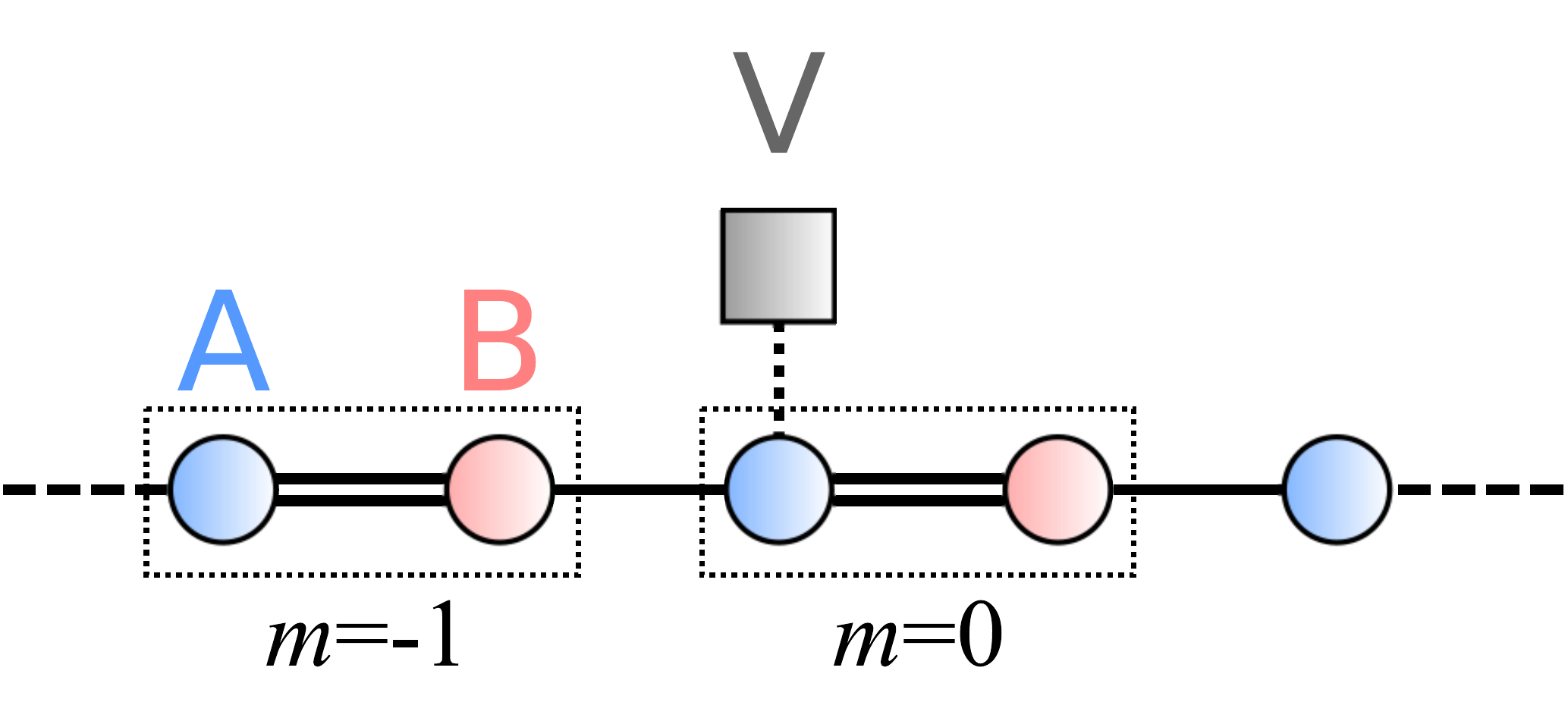}\\
\includegraphics[trim = 05mm 0mm 9mm 0mm, clip, width=4.2cm]{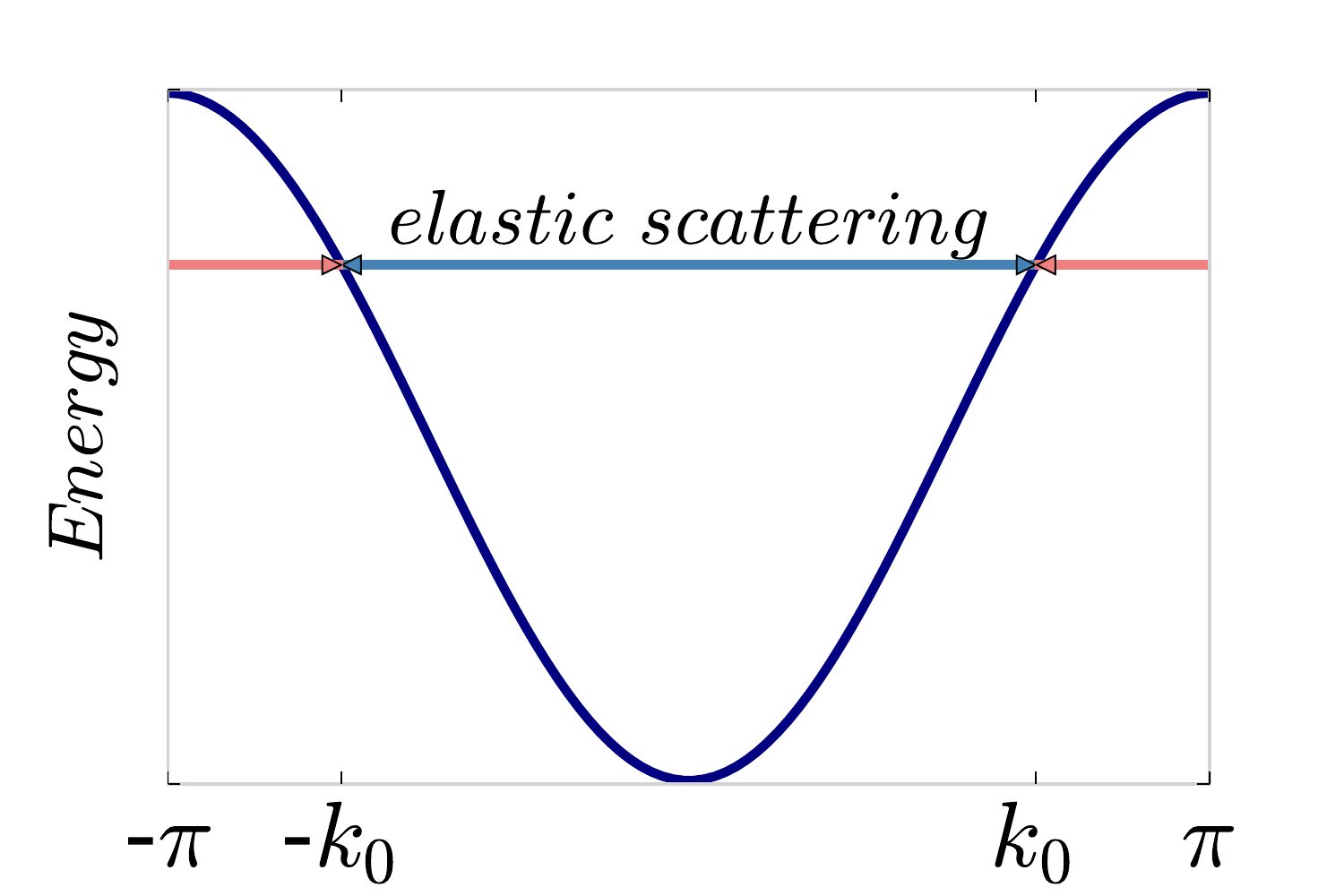}&
\includegraphics[trim = 05mm 0mm 9mm 0mm, clip, width=4.2cm]{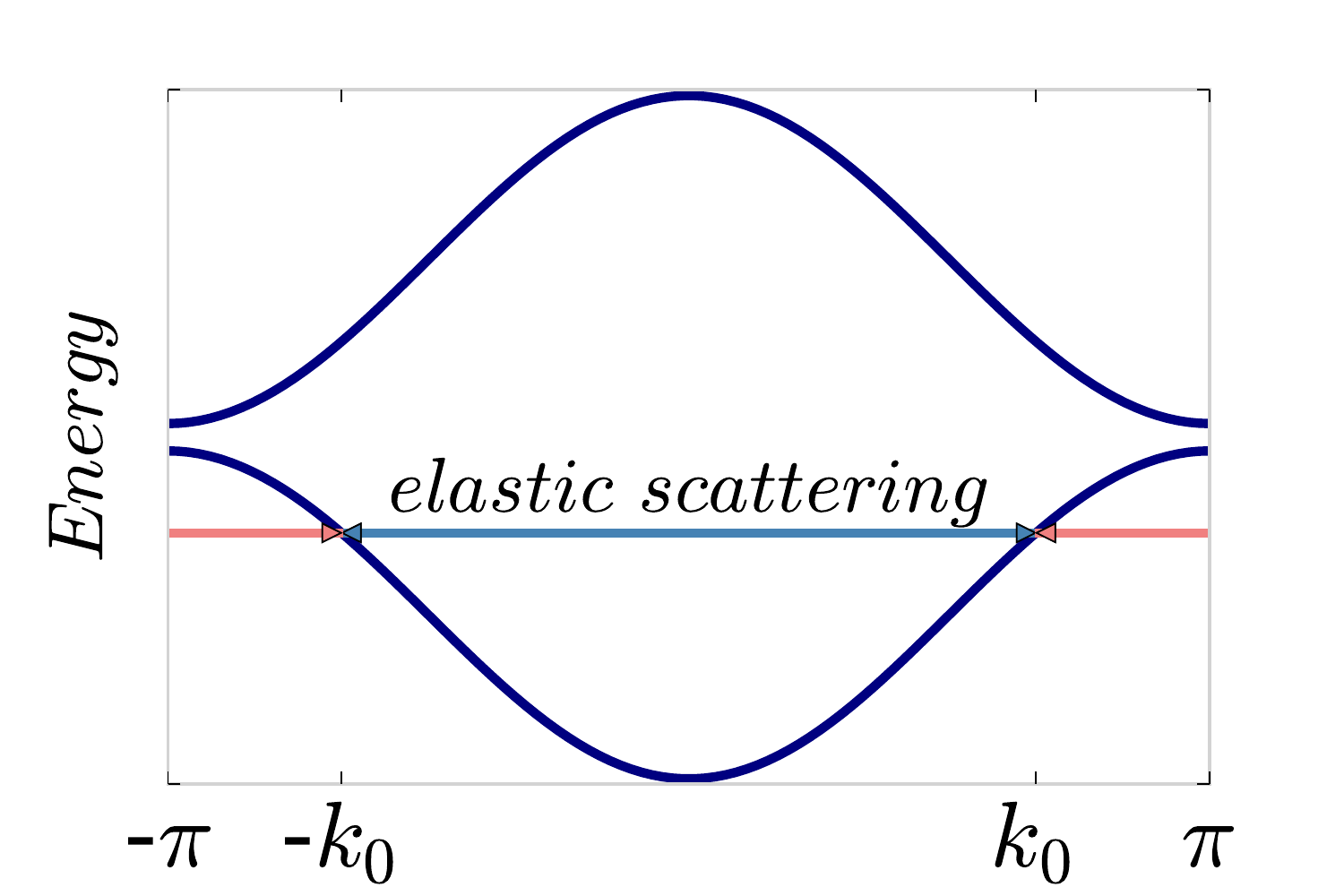}
\end{array}$
\caption{\small (Color online) Top: monatomic (left) and diatomic (right) pattern crystals with a localized impurity $V$. Bottom: Bloch spectra of the monatomic (left) and diatomic (right) pattern crystals in the first Brillouin zone. The horizontal double arrows depict elastic scattering processes between time-reversed states $-k_{0}$ and $k_{0}$.}
\label{Spectra}
\end{figure}

Electric screening in metals arises as a collective response of the conduction electrons to the Coulomb potential of a charged impurity. A long wavelength description of the problem captures the exponential screening of the impurity by the surrounding electrons, in agreement with the classical picture that depicts electrons as point charges. In quantum mechanics, however, particles are described by wavefunctions and may interfere. In the 1950s, J. Friedel actually understood that a charged impurity additionally yields a long-range interference pattern in the electronic density \cite{friedel1952xiv}. It consists of algebraically decaying $2k_{F}$-wavevector oscillations, and results from Fermi surface nesting associated to twice the Fermi momentum $k_{F}$. Thus, these oscillations reported by Friedel for charges in metals rely on wave features, and they have subsequently been revisited in other contexts, such as magnetic interactions \cite{PhysRev.96.99,kasuya1956theory,yosida1957magnetic,fert1980role} and noninteracting electrons \cite{adhikari1986quantum,barlette2000quantum,dutreix2016friedel}. In particular, Friedel oscillations have been observed in nonrelativistic electron gases via scanning tunnelling microscopy (STM); an experimental technique that images the local density of states (LDOS), i.e., the electronic density with atomic-scale and energy resolutions \cite{crommie1993imaging,Sprunger21031997,PhysRevB.57.R6858}. These experiments have confirmed that backscattering was the most efficient process involved in the elastic scattering through short-range impurities. For noninteracting electrons in a one-dimensional crystal, backscattering is indeed responsible for $2k_{0}$-wavevector Friedel oscillations that behave as
\begin{align}\label{LDOS Monatomic}
\delta \rho (m,\omega) = V(\omega) \cos(2k_{0}m) \,.
\end{align}
Here $\delta \rho$ denotes the correction to the LDOS induced by a localized impurity, $m$ labels the distance to the impurity in units of the Bravais lattice vector, $V(\omega)$ is some real function, and wavevector $2k_{0}$ refers to the elastic backscattering between the time-reversed states $-k_{0}$ and $k_{0}$ at energy $\omega$ ($\hbar=1$). Such backscattering wavevectors are illustrated in Fig.\,\ref{Spectra} by the double arrows. The reader may find the details of the derivation of Eq.\,(\ref{LDOS Monatomic}) in Supplemental Material (SM) \cite{SM}.

In this letter, we would like to highlight the effects of internal degrees of freedom on the long-range interference pattern induced by a localized impurity. These degrees of freedom could for example consist of spins up and down, or charge-conjugated particles such as electrons and holes. Here, however, we rather consider they result from a sublattice structure, as illustrated in the right-hand column of Fig.\,\ref{Spectra}. This one-dimensional dimerized structure naturally appears in some organic compounds \cite{PhysRevLett.42.1698}, and may be viewed as resulting from the Peierls instability of a metallic crystal like the one depicted in the left-hand column of Fig.\,\ref{Spectra}. It has also been realized in acoustics, photonics and ultracold atomic gases \cite{atala2013direct,xiao2015geometric,zeuner2015observation,poli2015selective}. The system is obviously invariant under time and space inversions. For the sake of simplicity we restrict the discussion to two sublattices, namely, A and B, within a nearest-neighbor tight-binding approximation. Nonetheless, this is not detrimental to the relevance of the results presented below, and the reader may find a generalization to an arbitrary number of freedom degrees in SM \cite{SM}. The Bloch band structure is then characterized by a $2\times2$ Hermitian matrix which, in the sublattice basis $\{$A,$\,$B$\}$, generically reads $H(k)=\sum_{i=1}^{2}d_{i}(k)\,\sigma_{i}$, where $d_{1}(k)=t_{1}(\alpha + \cos k)$ and $d_{2}(k)=t_{1}\sin k$. The parameters $\alpha t_{1}$ and $t_{1}$ respectively denote the intra- and inter-dimer hopping amplitudes whose ratio is $\alpha$, while matrix $\sigma_{i}$ is the $i$th Pauli matrix. Since there is only one energy scale in this description, energy will be given in units of $t_{1}$, if not otherwise specified. This prototypical model is sometimes referred to as SSH model, with reference to the work of Su, Schrieffer, and Heeger about the formation of topological solitons in polyacetylene $C_{n}H_{n}$ \cite{PhysRevLett.42.1698}. Note that there is no term scaling with $\sigma_{3}$ under time and space inversions. Furthermore, we disregard any contribution scaling with the identity matrix in the Bloch Hamiltonian matrix, such as next-nearest-neighbor processes. This would neither change the Bloch wavefunctions, nor the effects they are responsible for in the elastic scattering. The band structure is then entirely characterized by the set of the dispersion relation $\mathcal{E}_{\pm}(k)=\pm [d_{1}^{2}(k)+d_{2}^{2}(k)]^{1/2}$ and the Bloch eigenstates $\langle \pm,k| = (1,\,\pm e^{i\theta_{k}})$. The Bloch eigenstates are not gauge invariant, obviously, since $e^{i\phi_{k}}|\pm,k\rangle$ is also eigenstate for any arbitrary $\phi_{k}$. But function $\theta_{k}$, which defines the phase shift between the two internal degrees of freedom of Bloch spinors, is not affected by such a gauge change. However, this phase shift turns out to be ill-defined, because there exists an ambiguity when defining the diatomic unit cell of the translationally invariant crystal. Indeed, this phase shift can be defined either as $\theta_{k}$, or as $\theta_{k}-k$. Of course, the convention we chose to describe the system will never affect the observable we will be focussing on, namely, the LDOS \cite{SM}.

\begin{figure}[b]
\centering
\includegraphics[trim = 5mm 5mm 2mm 5mm, clip, width=8.2cm]{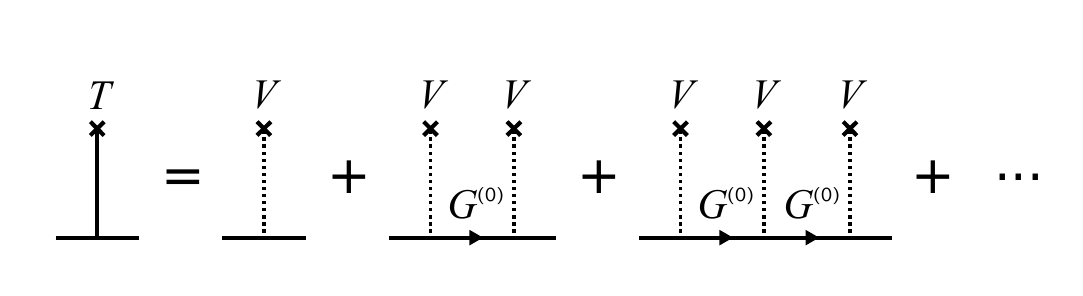}
\caption{\small Diagrammatic representation of the $T$-matrix within a perturbation theory in the impurity potential $V$. The oriented lines between two elastic scattering processes denote the bare Green's functions in momentum space.}
\label{Diagrams}
\end{figure}

\begin{figure}[t]
\centering
$\begin{array}{c}
\includegraphics[trim = 20mm 0mm 20mm 0mm, clip, width=6.0cm]{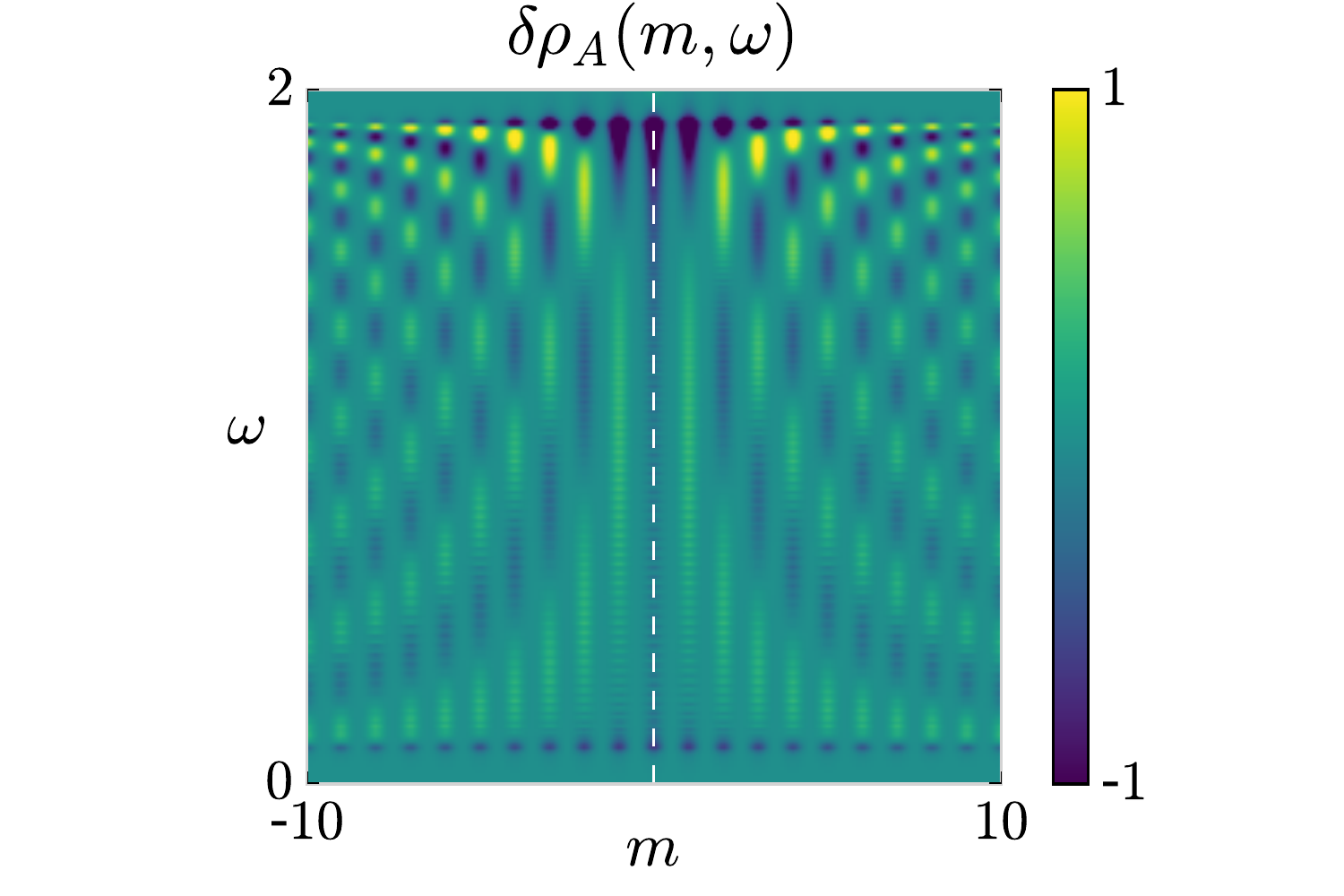}\\
~\\
\includegraphics[trim = 20mm 0mm 20mm 0mm, clip, width=6.0cm]{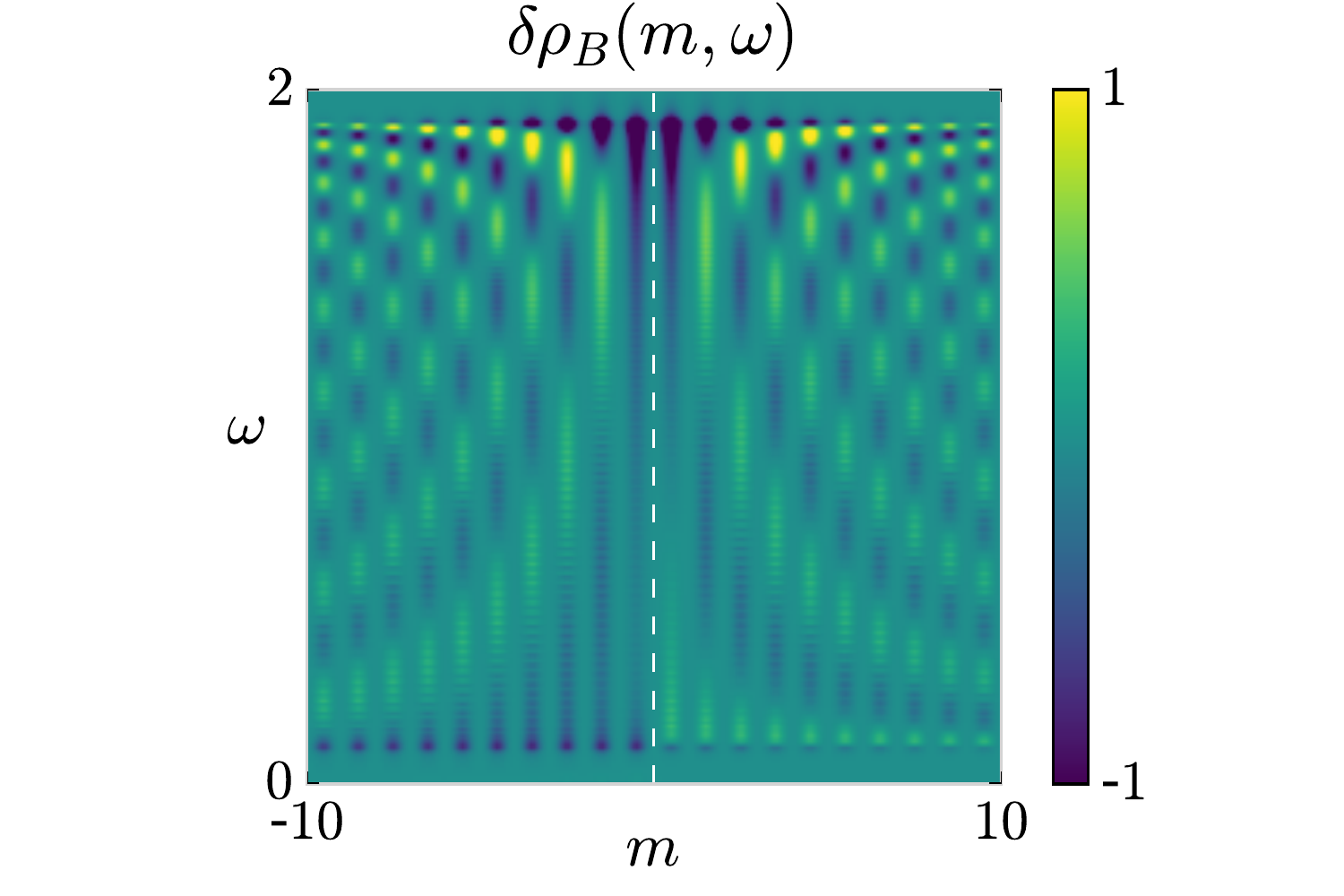}
\end{array}$
\caption{\small (Color online) Energy-resolved interference pattern induced in the LDOS of sublattices A (top) and B (bottom) by a localized impurity. The latter, which is simulated by a potential of $V=1$, belongs to sublattice A and unit cell $m=0$. Its position is marked by the vertical white dashed line. Energy is given in units of hopping $t_{1}$.}
\label{LDOS}
\end{figure}

From now on, we consider a localized impurity on sublattice A, which is simulated by $V\,\delta_{m,0}$. The impurity breaks the translational invariance of the crystal and defines a natural origin. The elastic scattering experienced by Bloch wavefunctions on the defect is described within a $T$-matrix approach. This consists of a perturbation theory in the impurity potential $V$. As shown in Fig\,\ref{Diagrams}, for such a localized potential the scattering diagrams define a geometric series and all the orders can be summed up exactly, regardless of the magnitude of the impurity potential. After introducing the retarded bare Green's matrix as $G^{(0)}(k,\omega) = [\omega+i\epsilon-H(k)]^{-1}$, where $\epsilon^{-1}>0$ describes a finite quasiparticle lifetime, the $T$-matrix reads
\begin{align}
T(\omega)=t(\omega)
\left(\begin{array}{cc}
1 & 0 \\
0 & 0 \\
\end{array}\right)
\,.
\end{align}
Here $t(\omega)=V\,[1-V\int dk\, G^{(0)}_{AA}(k,\omega)]^{-1}$ does not depend on wavevector $k$,  due to the integration that runs over the Brillouin zone. The correction to the LDOS in the presence of the impurity is finally obtained as
\begin{align}\label{LDOS Real Space}
\delta\rho(m,\omega)=-\frac{1}{\pi} \Imag \Tr \left[ G^{(0)}(m,\omega)\, T(\omega)\, G^{(0)}(-m,\omega) \right] \,
\end{align}
in the limit $\epsilon \rightarrow 0^{+}$. According to Fig.\,\ref{Spectra}, the dispersion relation of the valence band looks the same as the one of the monatomic crystal, and so do the elastic scattering wavevectors between two time-reversed states $-k_{0}$ and $k_{0}$. Therefore, if one restricts the elastic scattering problem to a spectral analysis, one would naively expect the localized impurity to induce Friedel oscillations that behave as in Eq.\,(\ref{LDOS Monatomic}) on both sublattices, A and B. The interference patterns obtained from the numerical evaluation of Eq.\,(\ref{LDOS Real Space}) for $V=1$ and $\alpha=0.9$ are shown in Fig.\,\ref{LDOS}. These patterns are resolved in energy, and the latter is restricted to positive values because we arbitrarily focus on the conduction band ($|1-\alpha|\leq\omega\leq1+\alpha$). Whereas the interferences on sublattice A are symmetric with respect to the impurity, they turn out to be asymmetric on sublattice B at low energies. This asymmetry is clearly in disagreement with the behavior of the Friedel oscillations introduced in Eq.\,(\ref{LDOS Monatomic}), and it cannot simply be understood from spectral features.

In order to get some insight into this asymmetric interference pattern, we perform the T-matrix approach analytically~\cite{SM}. The LDOS corrections it leads to for the two sublattices read
\begin{align}\label{LDOS AB}
\delta \rho_{A} (m,\omega) &= V(\omega) \cos(2k_{0}m) \notag \\
\delta \rho_{B} (m,\omega) &= V(\omega) \cos(2k_{0}m+\varphi_{k_{0}}) \,,
\end{align}
where $m$ labels the diatomic unit cells from the impurity, $k_{0}$ is a pole of $G^{(0)}(k,\omega)$ defined by $\omega^{2}=\mathcal{E}_{\pm}^{2}(k_{0})$ for $-\pi \leq k_{0} \leq 0$. Thus, if the spectral analysis illustrated in Fig.\,\ref{Spectra} is sufficient to understand the $2k_{0}$-wavevector oscillations as resulting from elastic backscattering between time-reversed states $k_{0}$ and $-k_{0}$, it does not explain the existence of the phase shift in the Friedel oscillations on the pristine sublattice, namely, sublattice B, which yields the asymmetric interferences shown in Fig.\,\ref{LDOS}. It has to be stressed that this phase shift does not arise from the condition that the scattering wavefunctions have to satisfy at the boundary with the impurity, as initially introduced by Friedel for finite size defects \cite{friedel1952xiv}. In the case of a localized impurity, indeed, there is no Friedel phase shift, as already shown in Eq.\,(\ref{LDOS Monatomic}). The phase shift involved on sublattice B actually arises from the internal degrees of freedom of the Bloch wavefunctions involved in the elastic scattering. It explicitly reads
\begin{align}\label{Phase Shift}
\varphi_{k_{0}}=\theta_{k_{0}}-\theta_{-k_{0}}= 2\theta_{k_{0}} \,,
\end{align}
where the last equality may be understood as resulting from time reversal symmetry. It is directly related to $\theta_{k_{0}}$, that is, the phase shift between the two components of the Bloch spinor of state $k_{0}$. Importantly, the LDOS corrections on sublattices A and B are observables; they are for instance accessible in atomic-scale-resolved STM experiments. So they do not depend on the ambiguity there is in the definition of the diatomic unit cell. This issue is explicitly fixed in SM \cite{SM}.

\begin{figure}[t]
\centering
$\begin{array}{cc}
\includegraphics[trim = 13mm 0mm 11mm 0mm, clip, width=4.2cm]{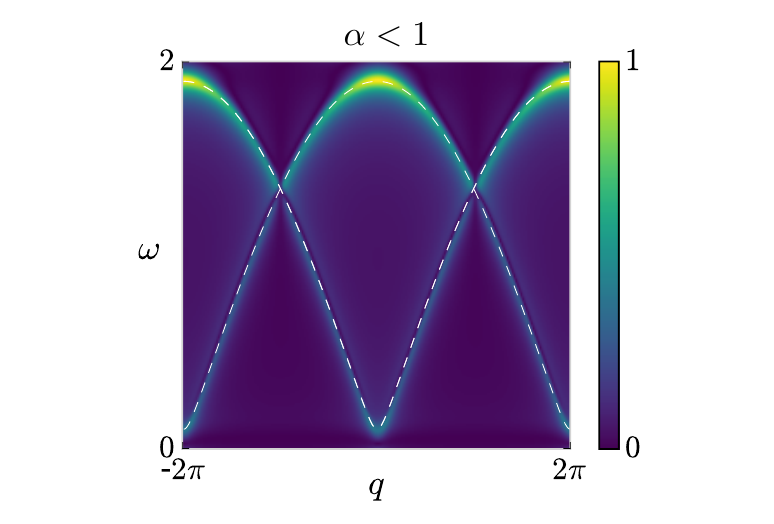}&
\includegraphics[trim = 13mm 0mm 11mm 0mm, clip, width=4.2cm]{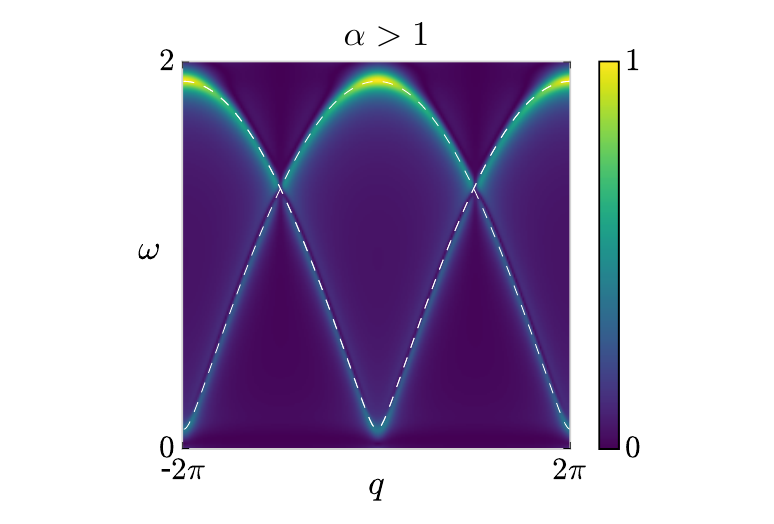}\\
\includegraphics[trim = 13mm 0mm 11mm 0mm, clip, width=4.2cm]{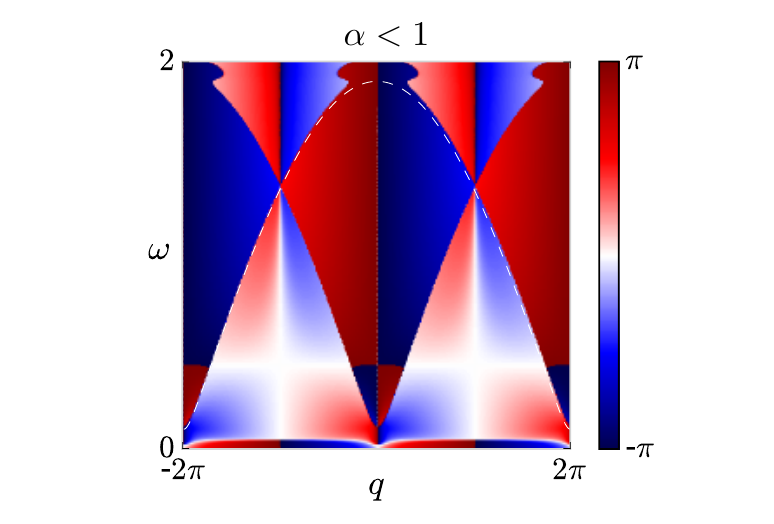}&
\includegraphics[trim = 13mm 0mm 11mm 0mm, clip, width=4.2cm]{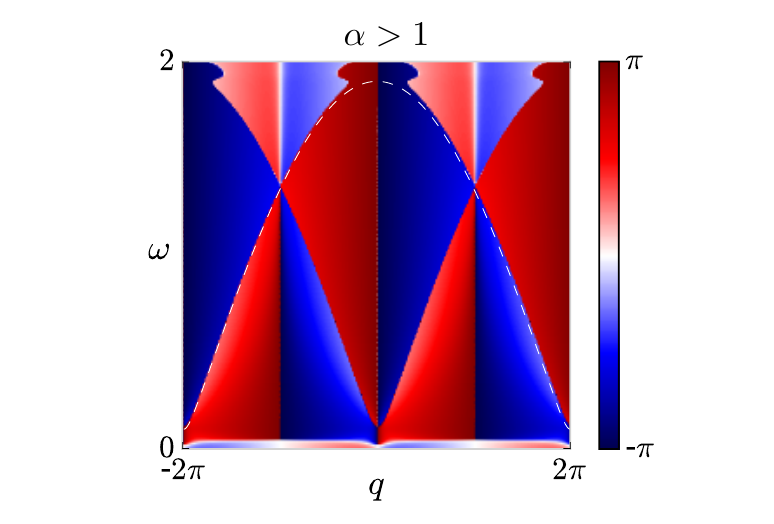}\end{array}$
\caption{\small (Color online) Modulus (top) and argument (bottom) of the Fourier transform of the energy-resolved interference pattern on sublattice B for $\alpha<1$ (left) and $\alpha>1$ (right). The white dashed lines mark the wavevectors $q=\pm2k_{0}$ at energy $\omega$.}
\label{kDOS}
\end{figure}

The Fourier analysis of interference patterns is often very instructive too. For example, it has demonstrated the ability of STM to probe the Fermi iso-energy contours in nonrelativistic electron gases \cite{Sprunger21031997,PhysRevB.57.R6858}, as well as the absence of backscattering of the massless relativistic charge carriers in graphene, and at the surfaces of three-dimensional topological insulators \cite{PhysRevLett.101.206802,roushan2009topological,xia2009observation}. Here, the Fourier transform of the Friedel oscillations introduced in Eq.\,(\ref{LDOS AB}) leads to Dirac combs
\begin{align}\label{kDOS AB}
\delta \rho_{A} (q,\omega) &= V(\omega)\, \sum_{n=-\infty}^{+\infty} \delta(q-2k_{0}+n2\pi)\notag \\
\delta \rho_{B} (q,\omega) &= V(\omega)\, \sum_{n=-\infty}^{+\infty} \delta(q-2k_{0}+n2\pi)\,e^{-i\varphi_{k_{0}}} \,,
\end{align}
where $\delta$ denotes the Dirac delta function, and $-\pi\leq k_{0}\leq +\pi$. Thus, the Fourier transform is a $4\pi$-periodic function of $q$ \cite{SM}. The modulus of the Fourier transform is the same for both sublattices, A and B. Its intensity is maximum for the backscattering wavevectors $q=2k_{0}\,[2\pi]$. This can be understood from the spectral features of the band structure in the absence of the impurity, as shown by the double arrows in Fig.\,\ref{Spectra}. This behavior is in agreement with the top panel of the left-hand column in Fig.\,\ref{kDOS}, which represents the Fourier transform of the interference pattern of sublattice B, previously depicted in Fig.\,\ref{LDOS}. It results from the numerical evaluation of the Fourier transform of the LDOS correction
\begin{align}\label{LDOS Momentum Space}
\delta\rho(q,\omega)=&\frac{i}{2\pi} \Tr \int dk \left[ G^{(0)}(k+q,\omega)\, T(\omega)\, G^{(0)}(k,\omega) \right] \notag \\
-&\frac{i}{2\pi} \Tr \int dk \left[ G^{(0)}(k,\omega)\, T(\omega)\, G^{(0)}(k+q,\omega) \right]^{*} \,
\end{align}
for the impurity potential $V=1$ and $\alpha=0.9$. The high intensity areas agree with the white dashed lines that mark the Dirac delta functions when the backscattering wavevectors $q=2k_{0}\,[2\pi]$ described by Eq.\,(\ref{kDOS AB}). Note that only the Fourier transform for sublattice B is shown in Fig.\,\ref{kDOS}, since the modulus of the Fourier transform for sublattice A is identical.

The argument of the Fourier transform along the maximum-intensity lines reveals the phase shift involved in the Friedel oscillations on sublattice B, according to Eq.\,(\ref{kDOS AB}). It is represented in the second plot of the left-hand column in Fig.\,\ref{kDOS}. Note that only one white dashed line is shown in the figure. It defines a periodic line in $(q,\omega)$-space that we denote by $\mathscr{C}_{2k_{0}}$. This line corresponds to the wavevectors $q=2k_{0}$ depicted by the blue double arrow in Fig.\,\ref{Spectra}. The Umklapp scattering processes associated to the red double arrow would lead to another white dashed line from which we would learn the same information, so it is not shown. Path $\mathscr{C}_{2k_{0}}$ is a closed path, along which we can define the following mapping:
\begin{align}
\varphi: q\in S^{1}=[-2\pi,\,2\pi] \mapsto \varphi_{q/2}\in S^{1}=[-\pi,\,\pi]
\end{align}
This mapping supports a topological characterization that indexes equivalence classes referred to as homotopy classes. They form some groups that are examples of topological invariants. As far as we are concerned, the topology of the mapping $\varphi: S^{1} \mapsto S^{1}$ is characterized by the first homotopy group of spheres, namely, $\pi_{1}(S^{1})=\mathbb{Z}$. This topological invariant is an integer that counts the number of times that $\varphi_{q/2}$ winds around the circle $S^{1}=[-\pi,\,\pi]$, when scattering wavevector $q$ runs once along $S^{1}=[-2\pi,\,2\pi]$. This is nothing but the winding number given by
\begin{align}
\mathcal{W} &= \frac{1}{2i\pi} \oint_{\mathscr{C}_{2k_{0}}} dq\, \partial q \ln \left[ \frac{\delta\rho_{A} (q,\omega)}{\delta\rho_{B} (q,\omega)}\right] = \int_{-\pi}^{\pi}\frac{dq}{2\pi}\, \partial_{q}\, \varphi_{q} \,.
\end{align}
From the second plot of the left-hand column in Fig.\,\ref{kDOS}, we infer that $\mathcal{W}=2$. When changing the ratio between intra- and inter-dimer hoppings to $\alpha>1$, one gets the right-hand column of Fig.\,\ref{kDOS}. It has been obtained for parameters $\alpha=10/9$ and $t_{1}=0.9$, so that the energy spectrum as well as the modulus of the Fourier transform remain unchanged. The phase of the Fourier transform along the maximum intensity closed path $\mathscr{C}_{2k_{0}}$ no longer winds and $\mathcal{W}=0$. Therefore, the presence of a localized impurity, which breaks the translational invariance, induces interference patterns that support nonequivalent topological characterizations, depending on the global properties of the geometrical phase shift involved in Friedel oscillations.

Moreover, the geometrical phase shift satisfies $\varphi_{k_{0}}=2\theta_{k_{0}}$ under time-reversal symmetry. So the topological properties of the interference-pattern Fourier transform are intrinsically related to the Bloch band structure topology, which is itself characterized by the winding number $W$:
\begin{align}
\mathcal{W} &= 2 \int_{-\pi}^{\pi} \frac{dk}{2\pi}\, \partial_{k}\, \theta_{k} \notag =2W \,.
\end{align}
Topological invariant $\mathcal{W}$ is then obviously connected to the Zak phase of the Bloch wavefunctions in the absence of impurity, namely, $\gamma=i\oint_{BZ}dk \langle \pm,\,k | \partial_{k} | \pm,\, k \rangle$, via the following relation: $\mathcal{W}=2\gamma/\pi$. The Zak phase is analogous to the Berry phase for one-dimensional systems \cite{zak1989berry} where the the role of the periodic external parameter is played by the quasi-momentum running over the 1D Brillouin zone. It refers here to the gauge-invariant geometrical phase picked up by the Bloch wavefunctions along the Brillouin zone. This is a crucial quantity involved in many fields of physics, among which charge pumping, electric polarization, orbital magnetism, symmetry-protected topological order and edge states \cite{thouless1983quantization,king1993theory,resta1994macroscopic,PhysRevLett.112.026402,schnyder2008classification,delplace2011zak}. It must be stressed that the value of such a phase depends on the choice of origin in the Fourier transform for sublattice B, or equivalently on the ambiguity there is in the definition of the unit cell. Indeed, since the definition of $\theta_{k}$ depends on the unit cell convention we choose, so does the winding number $W$ it leads to. Despite this ambiguity, the difference of winding numbers $W(\alpha>1)-W(\alpha<1)$ does not depend on this choice and is a well-defined quantity. This is actually a well-known issue, and it explains for example why the polarization in a crystal, which depends on the Zak phase of Bloch wavefunction, is only defined modulo a quantum of polarization \cite{king1993theory,resta1994macroscopic}. \textit{A fortiori}, only $\mathcal{W}(\alpha>1)-\mathcal{W}(\alpha<1)$ is well-defined and enables the distinction between the Fourrier transform of two topologically nonequivalent interference patterns, as the ones of Fig.\,\ref{kDOS}.

In summary, we have addressed the scattering problem of a localized impurity in a one-dimesional crystal for Bloch waves that possess internal degrees of freedom. While the impurity obviously yields Friedel oscillations in the LDOS associated to backscattering wavevector $q=2k_{0}$, we have shown that the internal freedom degrees of Bloch waves are responsible for a geometrical phase shift. The latter does not relate to the nature of the impurity and is then intrinsically different from the so-called Friedel phase shift. The Fourier transform of the energy-resolved interference pattern has revealed the momentum dependence of this geometrical phase shift, whose global properties enable us to discriminate two topologically nonequivalent interference patterns. Remarkably, these topological features are intrinsically connected to the Zak phase under-space and time-inversion symmetries, which characterizes the Bloch band structure topology in the absence of impurity. Measurements of the Zak phase have already been realized, for example through Bloch oscillations \cite{atala2013direct} and in non-Hermitian systems with losses \cite{rudner2009topological,zeuner2015observation}, but these prescriptions remain rather unsuitable for electronic compounds. Since the LDOS is a physical observable that is accessible in acoustic and photonic crystals, ultracold atomic gases, and electronic materials via STM, the interference pattern induced by a localized defect offers \textit{a priori} a joint route to image the band structure topology. Besides, this consists of a bulk measurement since it probes Bloch bands properties. Above, we have indeed focused on elastic scattering occurring in the conduction band of a one-dimensional insulator. A strong impurity potential $V\gg1$ would be responsible for a symmetry-protected zero-energy edge state exponentially localized on one or the other side of the defect, depending on the Zak phase of the Bloch wavefunctions. Because the prescription we propose to probe the Zak phase is resolved in energy, it should then be possible to probe both the topological properties of Bloch bands, as well as the existence of zero-energy edge states they lead to within the band gap. This would offer a unique opportunity to observe the bulk-edge correspondence in a single experiment.

\begin{acknowledgments}
The authors are very grateful to D. Carpentier for stimulating discussions. This work was supported by the French Agence Nationale de la Recherche (ANR) under grant TopoDyn (ANR-14-ACHN-0031).
\end{acknowledgments}

\bibliography{references}

\newpage
~
\newpage
\onecolumngrid

\section{Supplemental Material for ``Geometrical Phase Shift in Friedel Oscillations''}

\subsection{Friedel oscillations for $N=2$}
This SM section details the derivations of the Friedel oscillations induced in the local density of states, as introduced in Eq.\,(1) and Eq.\,(4) in the main text.

\subsubsection{Retarded bare Green's matrix}
Within a two-band nearest-neighbor tight-binding description, the off-diagonal component of the bare Green's matrix may be written as
\begin{align}\label{GF Definition}
G^{(0)}_{BA}(m, \omega) &= \int_{-\pi}^{+\pi} \frac{dk}{2\pi}
\frac{(\alpha+\,e^{ik})\,e^{ikm}}{\omega^{2}-\mathcal{E}_{\pm}^{2}(k)} \\
&= \int_{-\pi}^{+\pi} \frac{dk}{2\pi}
\frac{(\alpha+\,e^{ik})\,e^{ikm}}{\omega^{2}-|\alpha+e^{ik}|^{2}} \\
&= \oint_{\mathscr{C}}\frac{dz}{2i\pi} \frac{(\alpha+\,e^{ik})\,z^{m}}{\mathcal{P}(z)} \,,
\end{align}
where it is assumed that $m\geq0$, $z=e^{ik}$, $\mathcal{P}(z)=-\alpha\,z^{2}+(\omega^{2}-1-\alpha^{2})\,z - \alpha$, $\mathscr{C}$ denotes the unit circle, and energy is given in units of hopping amplitude $t_{1}$. As we are interested in probing the bulk energy bands, which implies $(1-\alpha)^{2}\leq \omega^{2} \leq (1+\alpha)^{2}$, there are two complex roots for polynomial $\mathcal{P}$, namely,
\begin{align}
z_{\pm} &= \frac{\omega^{2}-1-\alpha^{2}\pm i\sqrt{\left[(1+\alpha)^{2}-\omega^{2}\right]\left[\omega^{2}-(1-\alpha)^{2}\right]}}{2\alpha} \,.
\end{align}
These roots satisfy $|z_{\pm}|=1$. Nevertheless, we have to consider a finite quasiparticle lifetime, which is achieved by adding a small imaginary part $i\epsilon$ to the frequency $\omega$. Moreover, we assume $\epsilon>0$ to work with the retarded Green function. Thus, $|z_{+}|>1$ and $z_{-}$ is the only pole that remains inside the unit circle. It is simply denoted $z_{0}=e^{ik_{0}}$ from now on. Since $z_{0}$ has a negative imaginary part, this implies $-\pi< k_{0}<0$. Therefore, in the limit $\epsilon \rightarrow 0^{+}$ the bare Green's function is given by
\begin{align}
G^{(0)}_{BA}(m, \omega)
&=
-\frac{i}{2v_{k_{0}}} e^{ik_{0}m+i\theta_{k_{0}}} \,
\end{align}
where we have used that $\theta_{k_{0}}=\Arg[\alpha+e^{ik_{0}}]$, and $\omega^{2}=|\alpha+e^{ik_{0}}|^{2}$, so that $v_{k_{0}}=\frac{d\omega}{dk}|_{k_{0}}=-\alpha\sin(k_{0})\,\omega^{-1}$. For simplicity we focus on the conduction bands ($\omega\geq0$) from now on. Besides, we have to be careful with the sign of $m$ which has been assumed to be positive so far. When $m<0$, then we can use the variable change $z=e^{-ik}$, which straightforwardly leads to
\begin{align}
G^{(0)}_{BA}(\pm|m|, \omega)
&=
-\frac{i}{2v_{k_{0}}} e^{ik_{0}|m|+i\theta_{\pm k_{0}}} \,.
\end{align}
The other off-diagonal component of the bare Green's matrix is obtained in the same way. This results in
\begin{align}
G^{(0)}_{AB}(\pm|m|, \omega)
&=
-\frac{i}{2v_{k_{0}}} e^{ik_{0}|m|-i\theta_{\pm k_{0}}} \,.
\end{align}
When doing the substitution $\alpha+e^{ik} \leftrightarrow \omega$ in the numerator of Eq.\,(\ref{GF Definition}), we immediately obtain the real-space representation of the diagonal component of the bare Green's matrix, that is
\begin{align}
G^{(0)}_{AA}(\pm|m|, \omega)
&= G^{(0)}_{BB}(\pm|m|, \omega) = -\frac{i}{2v_{k_{0}}} e^{ik_{0}|m|} \,.
\end{align}
Note finally that this latter expression does not depend on wether there is one or several atoms per unit cell, so that the LDOS correction $\delta\rho_{A}(m,\omega)$ it leads to will be the same for the monatomic and diatomic pattern crystals.

\subsubsection{$T$-matrix appraoch}
\begin{figure}[b]
\centering
$\begin{array}{cc}
\includegraphics[trim = 20mm 0mm 20mm 0mm, clip, width=5cm]{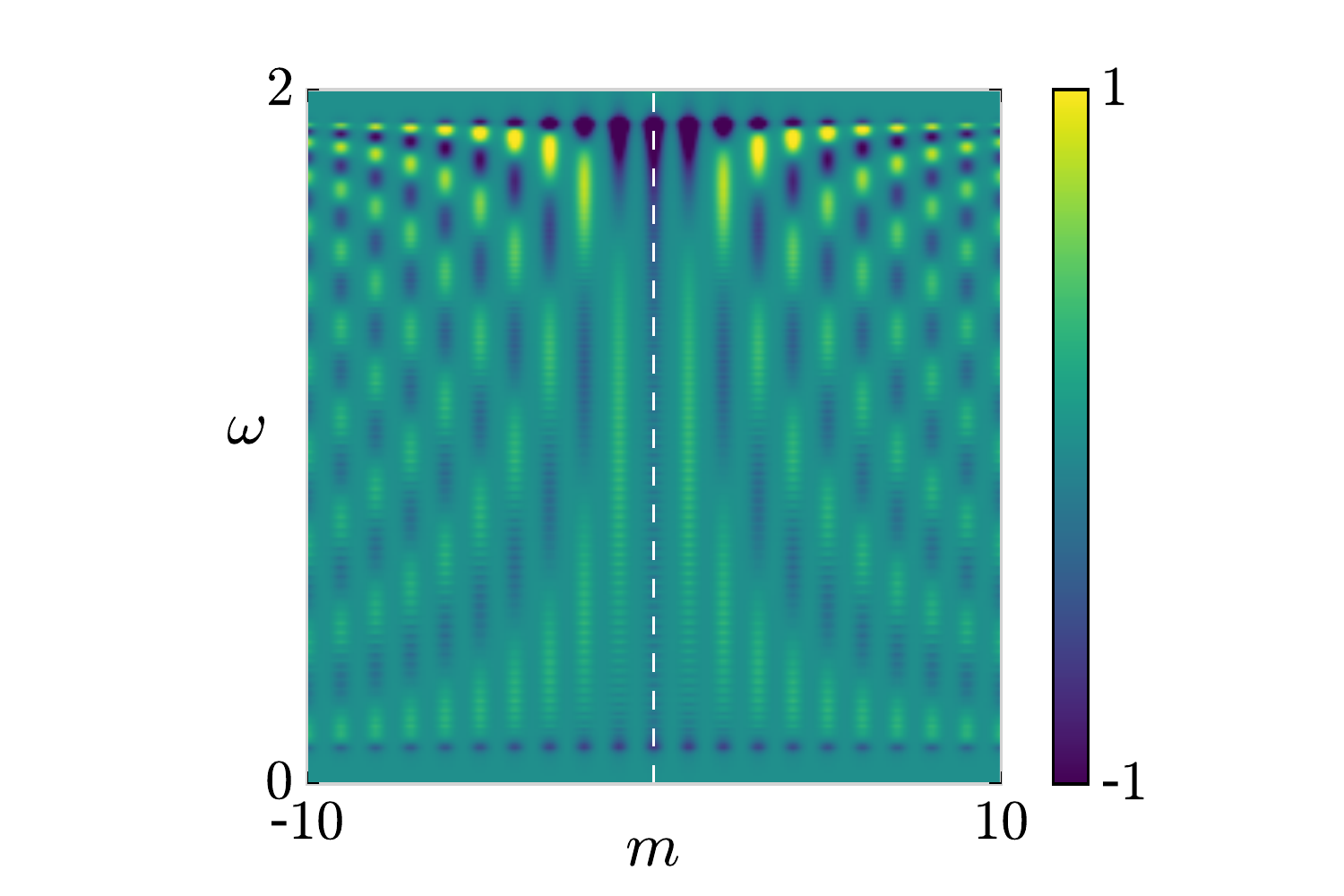}&
~~~~~~~~~~~~~\includegraphics[trim = 20mm 0mm 20mm 0mm, clip, width=5cm]{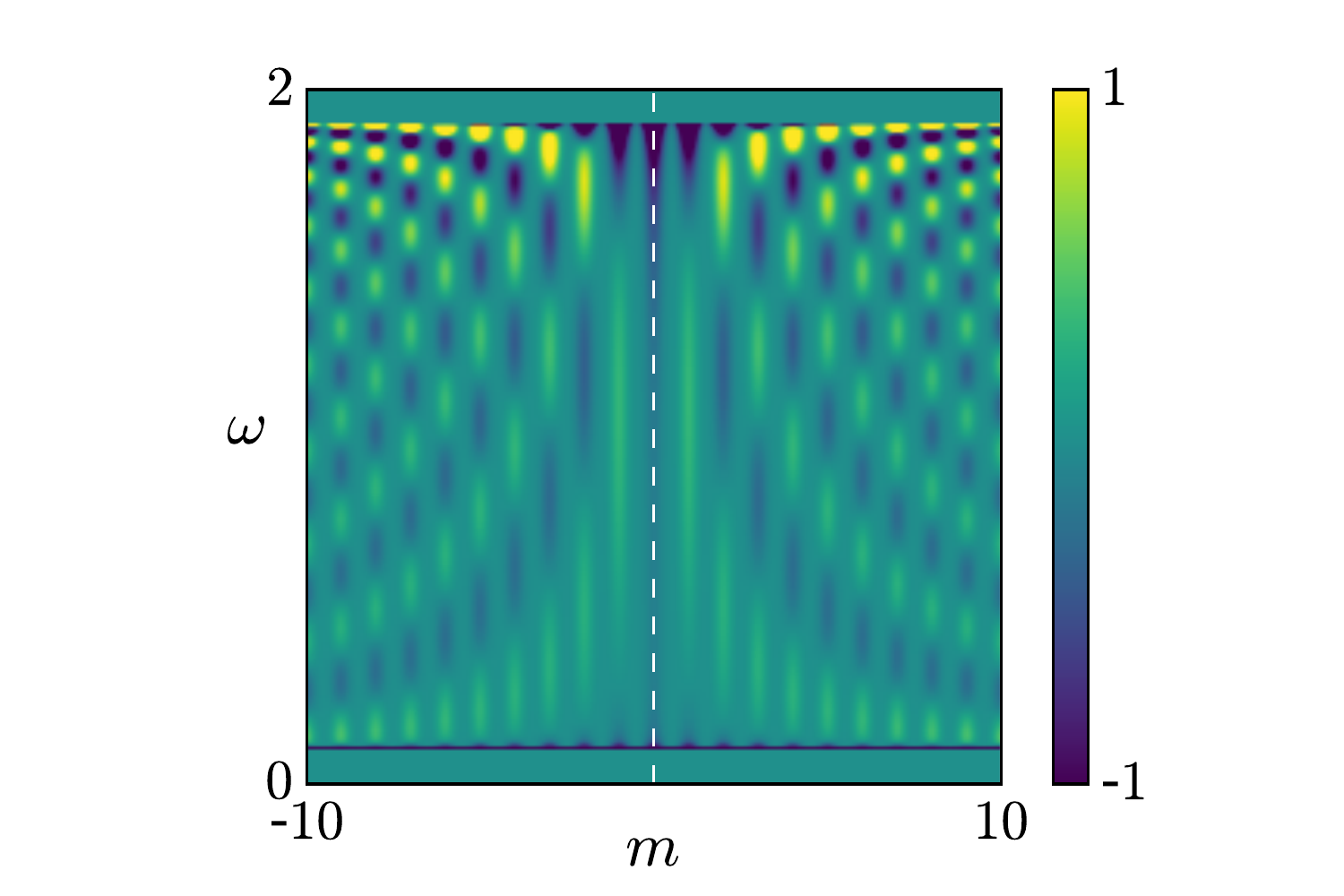}\\
\includegraphics[trim = 20mm 0mm 20mm 0mm, clip, width=5cm]{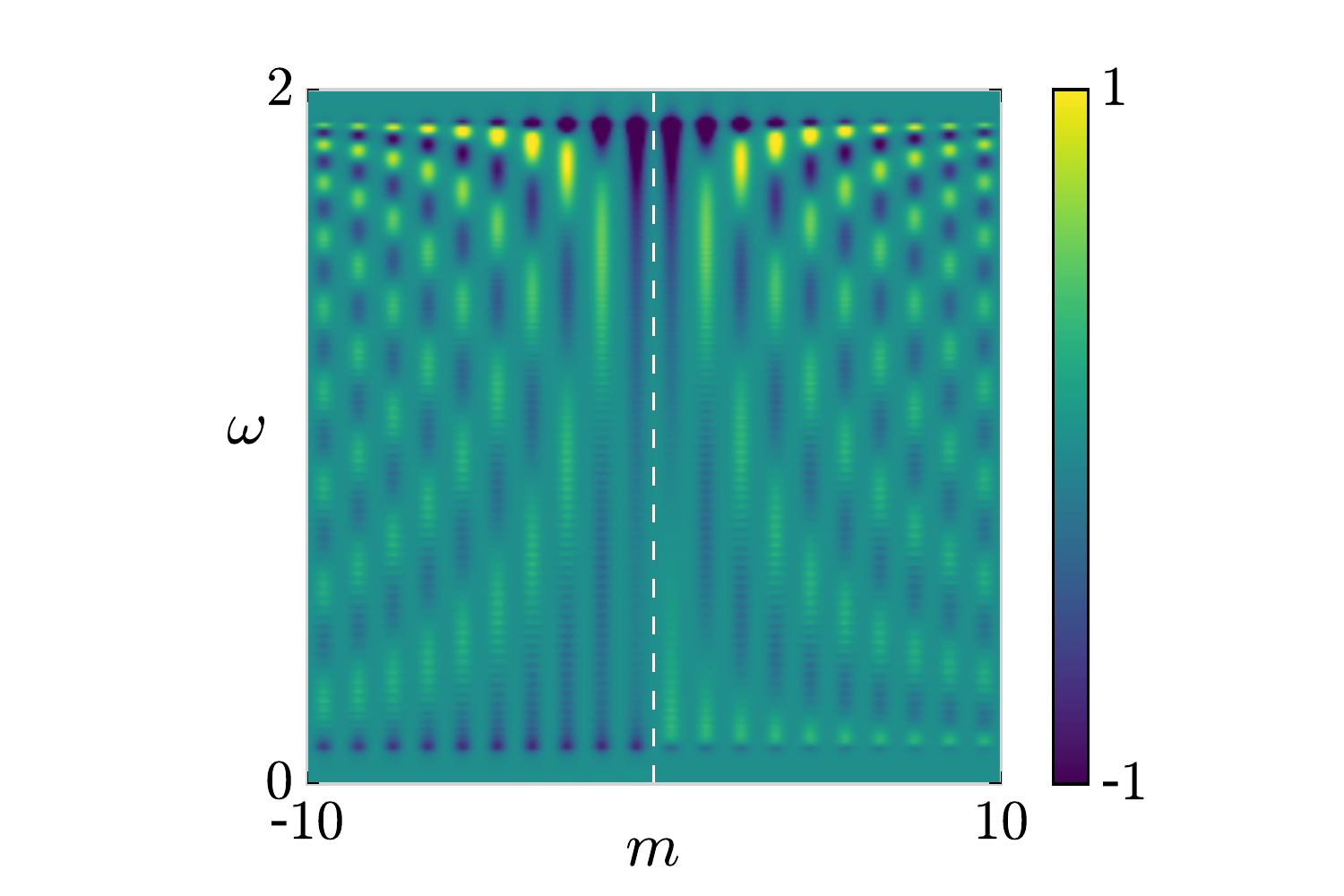}&
~~~~~~~~~~~~~\includegraphics[trim = 20mm 0mm 20mm 0mm, clip, width=5cm]{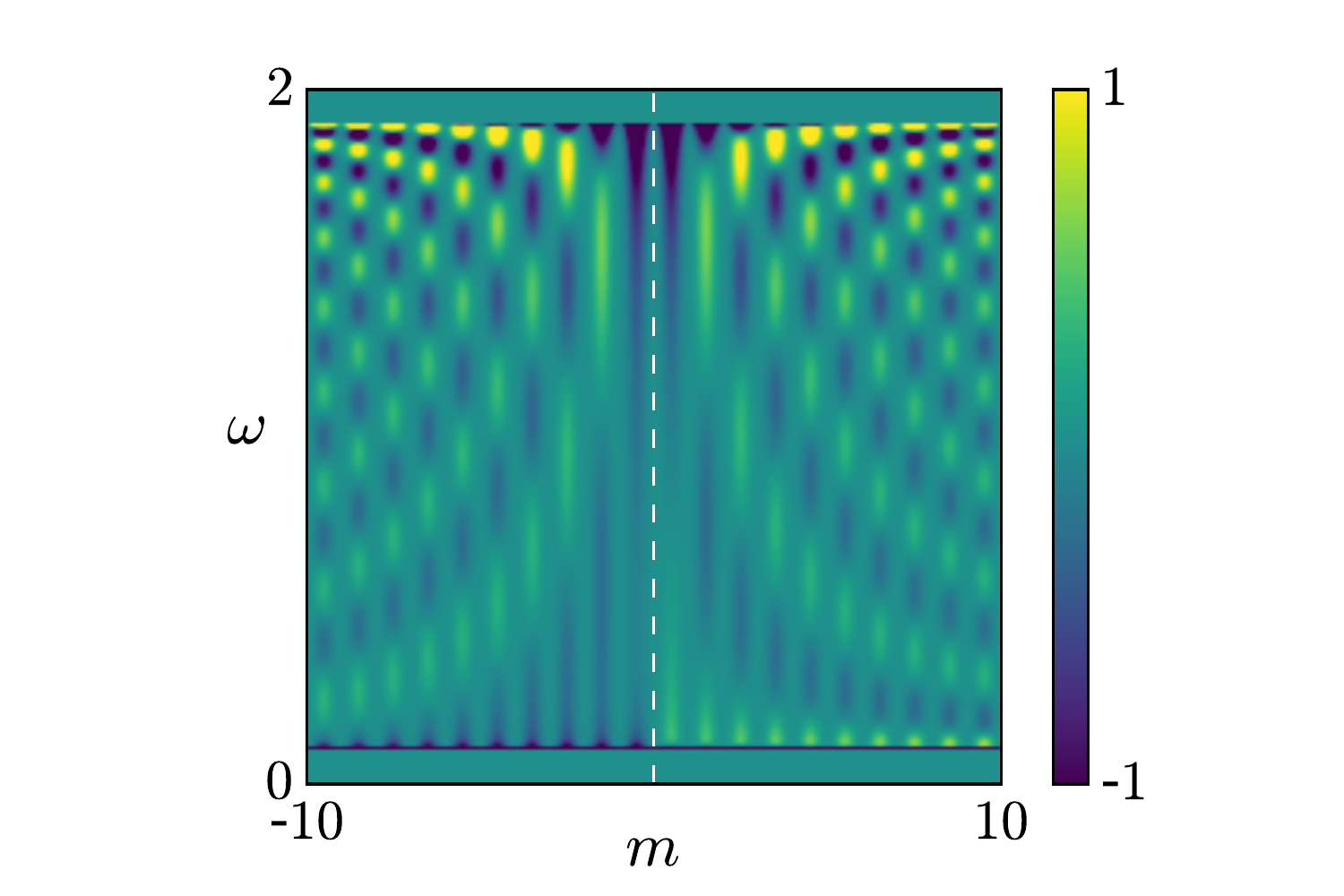}\\
\end{array}$
\caption{\small (Color online) Energy-resolved interference patterns in the LDOS on sublattice A (top) and sublattice B (bottom) obtained numerically (left) and analytically (right). Only the scattering occurring in the conduction band is shown, i.e., $|1-\alpha|\leq \omega \leq 1+\alpha$ where $\alpha=0.9$. The vertical white dashed line marks the site of the impurity that belongs to sublattice A and unit cell $m=0$. The impurity potential is simulated by $V=1$. Energy is given in units of the hopping amplitude $t_{1}$.}
\label{LDOS Check}
\end{figure}

The $T$-matrix satisfies
\begin{align}
T(\omega)=t(\omega)
\left(\begin{array}{cc}
1 & 0 \\
0 & 0 \\
\end{array}\right) \,,
\end{align}
where $t(\omega)=V\,[1-V\int dk\, G^{(0)}_{AA}(k,\omega)]^{-1}$ does not depend on wavevector $k$, due to the integration that runs over the Brillouin zone. From the real-space expression of the bare Green's functions introduced above, this can be rewritten as
\begin{align}\label{T-matrix component}
t(\omega) &= \frac{V}{1-VG^{(0)}_{AA}(0, \omega)} = -i\frac{2v_{k_{0}}V}{\sqrt{(2v_{k_{0}})^{2}+V^{2}}} e^{i\tau_{k_{0}}} \,,
\end{align}
where $\tau_{k_{0}}=\Arg[(V+i2v_{k_{0}})/\sqrt{(2v_{k_{0}})^{2}+V^{2}}]$.
The variation of local density of states on both sublattices is then obtained as the imaginary part of the following matrix product: $G^{(0)}(m,\omega)T(\omega)G^{(0)}(-m,\omega)$. It finally behaves as
\begin{align}\label{FO LDOS 0}
\delta\rho_{A}(m,\omega) &= V(\omega) \cos(2k_{0}m \pm \tau_{k_{0}}) \notag \\
\delta\rho_{B}(m,\omega) &= V(\omega) \cos(2k_{0}m \pm \tau_{k_{0}}+\varphi_{k_{0}}) \,.
\end{align}
where the sign ``$\pm$'' refers to the positive and negative values of $m$, $\varphi_{k_{0}}=\theta_{k_{0}}-\theta_{-k_{0}}$ and we have defined
\begin{align}
V(\omega)=\frac{1}{2v_{k_{0}}}\frac{V}{\sqrt{(2v_{k_{0}})^{2}+V^{2}}} \,.
\end{align}
As shown in Fig.\,\ref{LDOS Check} for $\alpha=0.9$ and $V=1$, these analytical expressions of Friedel oscillations are in agreement with the numerical evaluation of the LDOS correction. They describe interferences that are symmetric with respect to the impurity site on sublattice A, and asymmetric on sublattice B due to the phase shift $\varphi_{k_{0}}$.

Importantly, the T-matrix, which does not depend on space coordinates, has two contributions. On the one hand, it modulates the amplitudes of Friedel oscillations. On the other hand, it is responsible for the phase shift $\tau_{k_{0}}=\Arg[(V+i2v_{k_{0}})/\sqrt{(2v_{k_{0}})^{2}+V^{2}}]$. This phase shift is obviously bounded (i.e. can be chosen as monovalued), since $(V+i2v_{k_{0}})/\sqrt{(2v_{k_{0}})^{2}+V^{2}}$ cannot wind around the complex plane origin when varying $k_{0}$. As we are interested in the global geometrical properties of the phase shifts in Friedel oscillations, it is sufficient to consider that the LDOS correction is given by
\begin{align}\label{FO LDOS}
\delta\rho_{A}(m,\omega) &= \frac{1}{2v_{k_{0}}} \cos (2k_{0}m) \notag \\
\delta\rho_{B}(m,\omega) &= \frac{1}{2v_{k_{0}}} \cos (2k_{0}m + \varphi_{k_{0}})
\end{align}
in real space. These expressions actually correspond to the limits of strong impurity potential ($V\gg1$) or low energies ($k_{0}\sim\pm\pi$) in which $\tau_{k_{0}}$ vanishes. They already describe the $2k_{0}$-wavevector Friedel oscillations in the LDOS, as well as the global geometrical properties of the phase shift $\varphi_{k_{0}}$, which are the two universal behaviors that the main text aims to describe. Therefore, the expressions provided in Eq.\,(\ref{FO LDOS}) are the ones discussed in the main text. Note that they are discussed with respect to numerical evaluations of the LDOS correction for $V=1$ in the main text, which does not fall into the limits $V\gg1$ nor $k_{0}\sim\pm\pi$. Nonetheless, they show a very good agreement with each other, which confirms that the expressions in Eq.\,(\ref{FO LDOS}) already describes universal behaviors of Friedel oscillations.

\subsection{Ambiguity in the unit cell definition}
\begin{figure}[b]
\centering
\includegraphics[trim = 0mm 70mm 0mm 70mm, clip, width=6.cm]{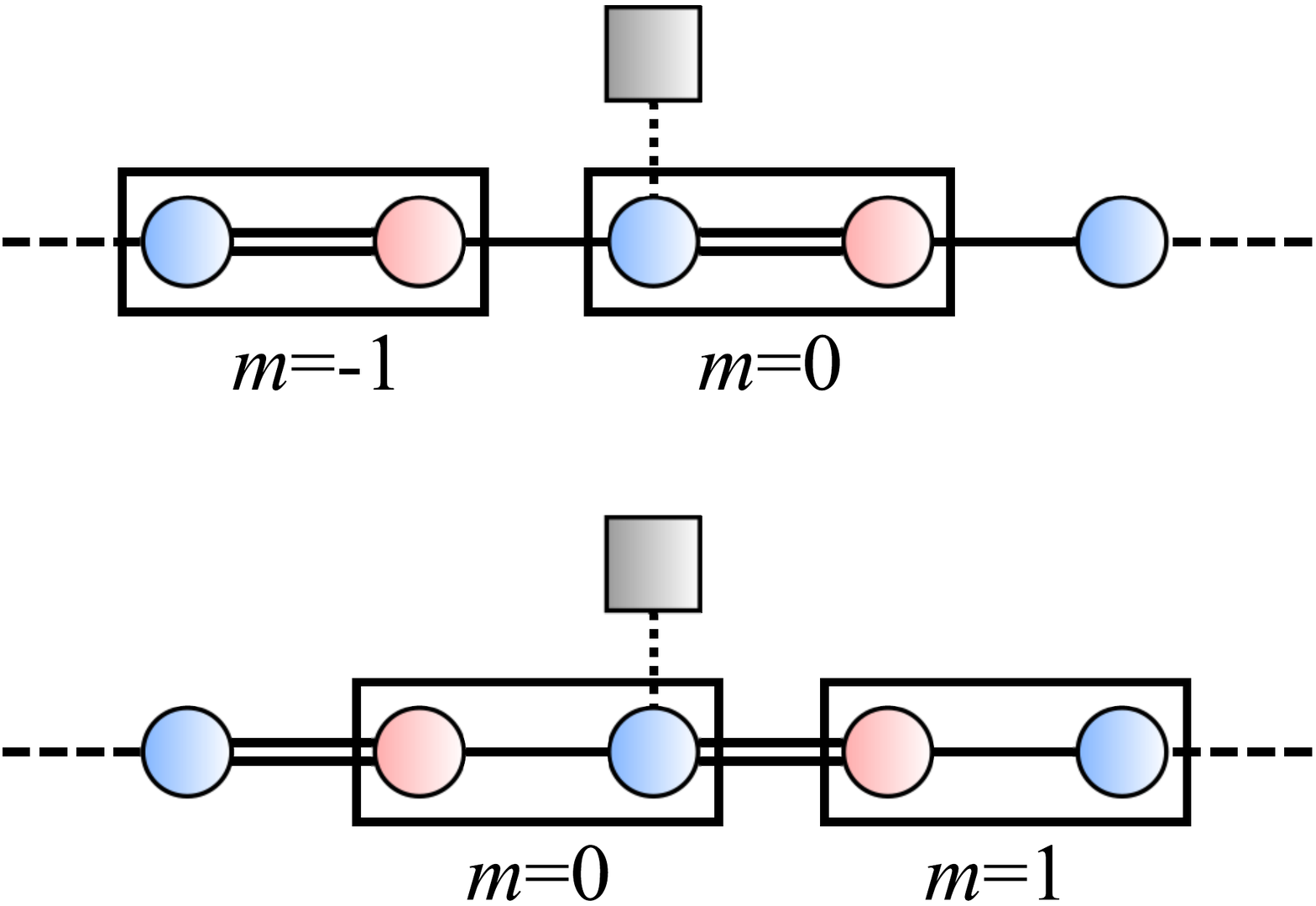}
\caption{\small (Color online) Illustration of the two possible definitions of the unit cell. The site of the impurity (grey square) unambiguously belongs to the site A (blue disk) of the unit cell $m=0$. Then, there are two nearest-neighbor sites B (red disks) and we have to pick one of them to obtain the diatomic pattern of the unit cell, which leads to the two configurations depicted in the figure.}
\label{UnitCell}
\end{figure}

Here, we would like to point out an ambiguity in the definition of the unit cell. Fig.\,\ref{UnitCell} depicts two copies of a dimerized crystal in the presence of a localized impurity. The latter fixes a natural origin, so that the impurity site unambiguously belongs to the unit cell $m=0$ of sublattice A. In order to define the diatomic pattern, we then have to chose one the two nearest-neighbor sites as belonging to the unit cell $m=0$. This choice leads to an ambiguity in the definition of the unit cell, hence the two configurations illustrated in the figure. Of course, the Friedel oscillations as introduced in Eq.\,(\ref{FO LDOS}) appear through an observable, namely, the LDOS, and they cannot depend on this choice of unit cell.

This can also be understood explicitly. The top and bottom configurations in Fig.\,\ref{UnitCell} respectively correspond to the following Bloch Hamiltonian matrices:
\begin{align}
H^{(1)}(k) =
\left(\begin{array}{cc}
0 & \alpha + e^{-ik} \\
\alpha + e^{ik} & 0 \\
\end{array}\right)
~~~~~\text{and}~~~~~
H^{(2)}(k) =
\left(\begin{array}{cc}
0 & e^{ik}(\alpha + e^{-ik}) \\
e^{-ik}(\alpha + e^{ik}) & 0 \\
\end{array}\right)
 \,.
\end{align}
After reproducing the derivations of the bare Green's functions introduced above, we end up with the following expressions for the Friedel oscillations on sublattice B:
\begin{align}
\delta\rho_{B}^{(1)}(\pm|m|,\omega) \propto \cos (2k_{0}|m| \pm 2\theta^{(1)}_{k_{0}})
~~~~~\text{and}~~~~~
\delta\rho_{B}^{(2)}(\pm|m|,\omega) \propto \cos (2k_{0}|m| \pm 2\theta^{(2)}_{k_{0}}) \,,
\end{align}
where $\theta^{(1)}_{k_{0}}=\Arg \left[\alpha+\,e^{ik_{0}}\right]$ and $\theta^{(2)}_{k_{0}}=\Arg \left[e^{-ik_{0}}(\alpha + e^{ik_{0}})\right]$. The two geometrical phase shifts are then related to one another via $\theta^{(2)}_{k_{0}}=\theta^{(1)}_{k_{0}}-k_{0}$, so that the LDOS correction on sublattice B can be rewritten as
\begin{align}
\delta\rho_{B}^{(1)}(\pm|m|,\omega) \propto \cos (2k_{0}|m| \pm 2\theta^{(1)}_{k_{0}})
~~~~~\text{and}~~~~~
\delta\rho_{B}^{(2)}(\pm|m|,\omega) \propto \cos (2k_{0}(|m|\mp1) \pm 2\theta^{(1)}_{k_{0}}) \,.
\end{align}
As it can be seen from these expressions, the Friedel oscillations in the LDOS are the same on every physical site and does not depend on the choice we made to label the unit cell.
For example, the site B of unit cell $m=0$ in the top configuration of Fig.\,\ref{UnitCell}, corresponds to the site B of unit cell $m=1$ in the bottom configuration of Fig.\,\ref{UnitCell}. The LDOS correction associated to this site is unambiguously given by
\begin{align}
\delta\rho_{B}^{(1)}(0,\omega) = \delta\rho_{B}^{(2)}(+1,\omega) \propto \cos (2\theta^{(1)}_{k_{0}}) \,.
\end{align}
Besides, the site B of unit cell $m=-1$ in the top configuration of Fig.\,\ref{UnitCell}, corresponds to the site B of unit cell $m=0$ in the bottom configuration of Fig.\,\ref{UnitCell}. The LDOS correction associated to this site is unambiguously given by
\begin{align}
\delta\rho_{B}^{(1)}(-1,\omega) = \delta\rho_{B}^{(2)}(0,\omega) \propto \cos (2\theta^{(2)}_{k_{0}}) \,.
\end{align}
Consequently, the interference pattern we describe, namely the Friedel oscillations in the observable LDOS, does not depend on the way we define the unit cell.

\subsection{Fourier transform of Friedel oscillations}
The Friedel oscillations on sublattice B described by the general expression in Eq.\,(\ref{FO LDOS 0}) satisfies
\begin{align}
\delta \rho_{B}(q,\omega)
=&V(\omega)  \sum_{n=-\infty}^{+\infty} \sum_{m>0} e^{i(2k_{0}+q+n2\pi)m}\, e^{i\varphi_{k_{0}}}\, e^{i\tau_{k_{0}}} + e^{-i(2k_{0}-q+n2\pi)m}\, e^{-i\varphi_{k_{0}}}\, e^{-i\tau_{k_{0}}}
\notag \\
+&V(\omega)  \sum_{n=-\infty}^{+\infty} \sum_{m<0} e^{i(2k_{0}+q+n2\pi)m}\, e^{i\varphi_{k_{0}}}\, e^{-i\tau_{k_{0}}} + e^{-i(2k_{0}-q+n2\pi)m}\, e^{-i\varphi_{k_{0}}}\, e^{i\tau_{k_{0}}}
\notag \\
=&V(\omega)  \sum_{n=-\infty}^{+\infty} \sum_{m>0} e^{i(2k_{0}+q+n2\pi)m}\, e^{i\varphi_{k_{0}}}\, e^{i\tau_{k_{0}}} + e^{-i(2k_{0}-q+n2\pi)m}\, e^{-i\varphi_{k_{0}}}\, e^{-i\tau_{k_{0}}}
\notag \\
+&V(\omega)  \sum_{n=-\infty}^{+\infty} \sum_{m>0} e^{-i(2k_{0}+q+n2\pi)m}\, e^{i\varphi_{k_{0}}}\, e^{-i\tau_{k_{0}}} + e^{+i(2k_{0}-q+n2\pi)m}\, e^{-i\varphi_{k_{0}}}\, e^{i\tau_{k_{0}}}
\notag \\
=&V(\omega)  \sum_{n=-\infty}^{+\infty} \frac{1}{1-e^{i(2k_{0}+q+n2\pi)}}\, e^{i\varphi_{k_{0}}}\, e^{i\tau_{k_{0}}} + \frac{1}{1-e^{-i(2k_{0}-q+n2\pi)}}\, e^{-i\varphi_{k_{0}}}\, e^{-i\tau_{k_{0}}}
\notag \\
+&V(\omega)  \sum_{n=-\infty}^{+\infty} \frac{1}{1-e^{-i(2k_{0}+q+n2\pi)}}\, e^{i\varphi_{k_{0}}}\, e^{-i\tau_{k_{0}}} + \frac{1}{1-e^{i(2k_{0}-q+n2\pi)}}\, e^{-i\varphi_{k_{0}}}\, e^{i\tau_{k_{0}}}
\notag \\
=&V(\omega)  \sum_{n=-\infty}^{+\infty} \frac{1-e^{-i(2k_{0}+q+n2\pi)}}{1-\cos(2k_{0}+q+n2\pi)}\, e^{i\varphi_{k_{0}}}\, e^{i\tau_{k_{0}}} + \frac{1-e^{i(2k_{0}-q+n2\pi)}}{1-\cos(2k_{0}-q+n2\pi)}\, e^{-i\varphi_{k_{0}}}\, e^{-i\tau_{k_{0}}+n2\pi}
\notag \\
+&V(\omega)  \sum_{n=-\infty}^{+\infty} \frac{1-e^{i(2k_{0}+q+n2\pi)}}{1-\cos(2k_{0}+q+n2\pi)}\, e^{i\varphi_{k_{0}}}\, e^{-i\tau_{k_{0}}} + \frac{1-e^{-i(2k_{0}-q+n2\pi)}}{1-\cos(2k_{0}-q+n2\pi)}\, e^{-i\varphi_{k_{0}}}\, e^{i\tau_{k_{0}}}
\notag \\
=&V(\omega)  \sum_{n=-\infty}^{+\infty} \frac{\cos(\tau_{k_{0}})-\cos(2k_{0}+q-\tau_{k_{0}}+n2\pi)}{1-\cos(2k_{0}+q+n2\pi)}\, e^{i\varphi_{k_{0}}} + \frac{\cos(\tau_{k_{0}})-\cos(2k_{0}-q-\tau_{k_{0}}+n2\pi)}{1-\cos(2k_{0}-q+n2\pi)}\, e^{-i\varphi_{k_{0}}}
\,,
\end{align}
where $-\pi\leq k_{0} \leq 0$, and we have used the fact that $z_{0}=e^{ik_{0}}$ lies inside the unit circle, before considering the limit $\epsilon \rightarrow 0$. Note that we have introduced a normalized summation over $n$, so that it satisfies $\sum_{n}=1$. Since $\tau_{-k_{0}}=-\tau_{k_{0}}$, this can be rewritten as
\begin{align}
\delta \rho_{B}(q,\omega)
=&V(\omega)  \sum_{n=-\infty}^{+\infty} \frac{\cos(\tau_{k_{0}})-\cos(2k_{0}-q-\tau_{k_{0}}+n2\pi)}{1-\cos(2k_{0}-q+n2\pi)}\, e^{-i\varphi_{k_{0}}}
\,,
\end{align}
where now $-\pi\leq k_{0} \leq \pi$ and $\delta \rho_{B}(q,\omega)$ is a $4\pi$-periodic function of $q$. For sublattice A, this straightforwardly leads to
\begin{align}\label{General FTLDOS}
\delta \rho_{A}(q,\omega)
=&V(\omega)  \sum_{n=-\infty}^{+\infty} \frac{\cos(\tau_{k_{0}})-\cos(2k_{0}-q-\tau_{k_{0}}+n2\pi)}{1-\cos(2k_{0}-q+n2\pi)}\,.
\end{align}

The Fourier transform of the Friedel oscillations introduced in Eq.\,(\ref{FO LDOS}) is given by
\begin{align}
\delta\rho_{A}(q,\omega) &= V(\omega) \sum_{n=-\infty}^{+\infty}[\delta(q+2k_{0}+n2\pi)\, + \delta(q-2k_{0}+n2\pi)] \notag \\
\delta\rho_{B}(q,\omega) &= V(\omega) \sum_{n=-\infty}^{+\infty}[\delta(q+2k_{0}+n2\pi)\,e^{+i\varphi_{k_{0}}} + \delta(q-2k_{0}+n2\pi)\,e^{-i\varphi_{k_{0}}}] \,,
\end{align}
where $-\pi \leq k_{0} \leq 0$. This subsequently reduces to
\begin{align}
\delta\rho_{A}(q,\omega) &= V(\omega) \sum_{n=-\infty}^{+\infty} \delta(q-2k_{0}+n2\pi) \notag \\
\delta\rho_{B}(q,\omega) &= V(\omega) \sum_{n=-\infty}^{+\infty} \delta(q-2k_{0}+n2\pi)\,e^{-i\varphi_{k_{0}}} \,.
\end{align}
where now $-\pi\leq k_{0}\leq\pi$, so that $-2\pi\leq q \leq2\pi$. Note once again that this behavior in reciprocal space agrees with the one described in Eq.\,(\ref{General FTLDOS}). The modulus is maximum for backscattering wavevectors $q=2k_{0}\,[2\pi]$, while the argument yields to the phase shift $\varphi_{k_{0}}$. The expression above is the one mentioned in the main text in Eq.\,(6).

\subsection{Generalization to $N$ internal degrees of freedom}

\subsubsection{Band structure under Chiral symmetry}
So far we have focused on the interference pattern induced in the LDOS when there are two sublattice degrees of freedom. It shows that, in the presence of space- and time-inversion symmetries, the interference pattern Fourier transform exhibits topological properties that are intrinsically related to the winding number characterizing the one-dimensional Bloch band structure. This relies on the space- and time-inversion symmetries, which prevents any mass term scaling with $\sigma_{3}$ in the two-band description. Since we disregard the processes scaling with the identity matrix $\sigma_{0}$, for they do not change the Bloch eigenstates, nor the topological properties they involve, our description relied on a bipartite Hamiltonian that had chiral symmetry. The Bloch band structure it is associated to then belongs to a chiral symmetry class, and the topological features it may exhibit are characterized by a winding number.

From now on, we consider a Bloch Hamiltonian matrix, namely $H(k)$, that satisfies a chiral symmetry: $\Gamma^{\dagger} H(k) \Gamma = -H(k)$, where the chiral operator $\Gamma$ is unitary and squares to plus identity. The chiral symmetry requires the eigenstates of $H(k)$, namely $|n(k)\rangle$, to come in pairs with opposite energies:
\begin{align}
H(k)\, |n(k)\rangle &= E_{n}(k)\, |n(k)\rangle \notag \\
H(k)\, \Gamma |n(k)\rangle &= -E_{n}(k)\, \Gamma |n(k)\rangle \,.
\end{align}
This obviously yields a particle-hole symmetric spectrum that we assumed to be gapped, i.e., $E_{n}(k)\neq0$ for all $k$ and $n$. Thus, there is an even number of energy bands, and we generically denote it $2N$, where $N$ is a natural number. Besides, we assume without loss of generality that the eigenstates $|n(k)\rangle$ (respectively $\Gamma |n(k)\rangle$) are normalized and associated to the negative (respectively positive) energy bands sorted in the ascending (respectively descending) order according to index $n \in [1\cdots N]$. Second, the chiral-conjugate eigenspaces are orthogonal to each other: $\langle n(k)| \Gamma | n(k)\rangle = 0$ for every $n$. Since $\Gamma^{2}=+1$, the chiral operator has two orthogonal subspaces of normalized states that have opposite chiralities:
\begin{align}\label{Basis Change}
|A_{n}(k)\rangle = \frac{|n(k)\rangle + \Gamma|n(k)\rangle}{\sqrt{2}}\, \text{~~~~~with~~~~~} \Gamma |A_{n}(k)\rangle = + |A_{n}(k)\rangle \,, \notag \\
|B_{n}(k)\rangle = \frac{|n(k)\rangle - \Gamma|n(k)\rangle}{\sqrt{2}}\, \text{~~~~~with~~~~~} \Gamma |B_{n}(k)\rangle = - |B_{n}(k)\rangle \,.
\end{align}
In particular, chiral operator $\Gamma$ supports a diagonal matrix representation of the form $\Gamma = \mathbb{I}_{N}\otimes \sigma_{3}$, where $\mathbb{I}_{N}$ denote the $N\times N$ identity matrix and $\sigma_{z}$ is the third Pauli matrix. For such a representation, the anticommutation relation that defines the chiral symmetry requires the Bloch Hamiltonian matrix to be off-diagonal:
\begin{align}
H(k) = 
\left( \begin{array}{cc} 
0 & F(k)  \\
F^{\dagger}(k) &  0
\end{array} \right) \,.
\end{align}
In the basis $\{ |A_{1}\rangle \cdots |A_{N}\rangle, |B_{1}\rangle \cdots |B_{N}\rangle \}$, the off-diagonal block $F(k)$ becomes diagonal, provided all energy bands $E_{n}(k)$ are non-degenerate. Otherwise it becomes trigonal, but it would not change the topological properties we aim to highlight here, since they relie on the assumption that there exists a gap around the zero-energy level. It reads
\begin{align}
F(k) = 
\left( \begin{array}{cccc} 
E_{1}(k)\,e^{-i\theta_{1}(k)} & 0 & \cdots & 0  \\
0 & \ddots & \ddots & \vdots \\
 \vdots & \ddots & \ddots & 0 \\
0 & \cdots & 0 & E_{N}(k)\,e^{-i\theta_{N}(k)}
\end{array} \right) \,.
\end{align}
In one dimension, the winding number that characterizes the first homotopy group of spheres of the Bloch band structure, namely $\Pi_{1}(S^{1})=\mathbb{Z}$, can be defined as
\begin{align}
W=\oint_{BZ} \frac{dk}{2i\pi} \nabla_{k} \ln \frac{\Det F^{\dagger}(k)}{|\Det F^{\dagger}(k)|} \,,
\end{align}
where
\begin{align}
\frac{\Det F^{\dagger}(k)}{|\Det F^{\dagger}(k)|}=\prod_{n=1}^{N} e^{i\theta_{n}(k)}=\exp{\left[ i\sum_{n=1}^{N}\theta_{n}(k)\right]} \,.
\end{align}
The expression of the winding number finally reduces to
\begin{align}
W=\sum_{n=1}^{N}W_{n} ~~~~~\text{where}~~~~~ W_{n}= \oint_{BZ} \frac{dk}{2\pi} \nabla_{k}\theta_{n}(k) \,.
\end{align}
It clearly appears through this expression that the winding number may become ill-defined if the assumption $E_{n}(k)\neq0$ is relaxed.

\subsubsection{Real-space representation of the bare Green's functions}
The retarded bare Green's function can then be introduced as
\begin{align}
G^{(0)}(k,\omega) 
= & [\omega - H(k) ]^{-1} \notag \\
= &\sum_{n=1}^{2N} \frac{|n(k)\rangle \langle n(k)|}{\omega - E_{n}(k)} \notag \\
= &\sum_{n=1}^{N} \left[ \frac{|n(k)\rangle \langle n(k)|}{\omega - E_{n}(k)} + \frac{\Gamma|n(k)\rangle \langle n(k)|\Gamma^{\dagger}}{\omega + E_{n}(k)} \right] \notag \\
= &\sum_{n=1}^{N} \left[ \frac{\omega}{\omega^{2}-E_{n}^{2}(k)} \left( |A_{n}\rangle \langle A_{n}| + |B_{n}\rangle \langle B_{n}| \right) + \frac{E_{n}(k)}{\omega^{2}-E_{n}^{2}(k)} \left( e^{i\theta_{n}(k)}\, |A_{n}\rangle \langle B_{n}| + e^{-i\theta_{n}(k)}\, |B_{n}\rangle \langle A_{n}| \right) \right] \,.
\end{align}
In the expression above, it has been used that $|A_{n}(k)\rangle = e^{i\theta_{n}^{A}(k)}\, |A_{n}\rangle$ and $|B_{n}(k)\rangle = e^{i\theta_{n}^{B}(k)}\, |B_{n}\rangle$, so that $\theta_{n}(k)=\theta_{n}^{A}(k)-\theta_{n}^{B}(k)$ is the phase shift between the freedom degrees of opposite chiralities for the $n$th energy band. Note that the chiral symmetry requires the Green's function to satisfy the following relation: $\Gamma^{\dagger} G^{(0)}(k,\omega) \Gamma = - G^{(0)}(k,-\omega)$. The matrix representation of the Green's function consists of uncoupled $2\times2$ blocks. Therefore, we can use the derivations already done in the previous sections of this Supplemental Material, and the Green's function can be written as
\begin{align}
G^{(0)}(m,\omega) 
= & e^{i2k_{0}m}\sum_{n=1}^{N} \frac{\delta(\omega \pm E_{n}(k_{0}))}{2iv_{n}(k_{0})} \left[ |A_{n}\rangle \langle A_{n}| + |B_{n}\rangle \langle B_{n}| + e^{i\theta_{n}(k_{0})}\, |A_{n}\rangle \langle B_{n}| + e^{-i\theta_{n}(k_{0})}\, |B_{n}\rangle \langle A_{n}| \right] \,,
\end{align}
where $m>0$, time-reversal symmetry requires the spectrum to satisfy $E_{n}(k)=E_{n}(-k)$, $v_{n}(k_{0})=\partial_{k}E_{n}(k_{0})$, and $k_{0}$ is assumed to be a pole satisfying $\omega=\pm E_{n}(k_{0})$ and $|e^{ik_{0}}|=1$. Of course, this pole depends on the band index $n$, but we omit it for more clarity. Here, a few remarks are in order about the Green's function expression above.

\begin{itemize}
\item The poles satisfying $|e^{ik_{\alpha}}|<1$ do not yield any long-range contribution, since they lead to exponential decays with the distance to the impurity, so they are disregarded.
\item The poles satisfying $|e^{ik_{\alpha}}|=1$ are assumed to be simple poles. They could be poles of higher orders, but it is always possible to make them simple pole by adiabatically varying the model parameters (e.g. hopping amplitudes), as long as the zero-energy gap does not close. This procedure does not change the winding number of the Bloch band structure $W$, nor the topological property of the geometrical phase shift we aim to highlight in the Friedel oscillations.
\item There may be several simple poles satisfying $|e^{ik_{\alpha}}|=1$. Nevertheless, we can still consider continuously deforming the spectrum by tuning adiabatically the model parameters without closing the zero-energy gap, in such a way that each energy band becomes a monotonic function of the momentum for $0\leq k \leq \pi$. This would bring us to a spectrum similar to the one discussed in the main text with only one simple pole $k_{0}$. 
\end{itemize}

\subsubsection{Friedel oscillations in the LDOS}
Let us consider a localized potential in real space such that it leads to the following representation of the T-matrix in momentum space:
\begin{align}
T(\omega) &= \sum_{n=1}^{N} t_{n}(\omega) \, |A_{n}\rangle \langle A_{n}| \,.
\end{align}
where 
\begin{align}
t_{n}(\omega) &= -i\frac{2v_{n}(k_{0})V_{n}}{\sqrt{(2v_{n}(k_{0}))^{2}+V_{n}^{2}}} e^{i\tau_{n}({k_{0}})} \,,
\end{align}
where $V_{n}$ is the potential experienced by the eigenstates of the $n$th band, and $\tau_{n}({k_{0})}=\Arg[(V_{n}+i2v_{n}(k_{0}))/\sqrt{(2v_{n}(k_{0}))^{2}+V_{n}^{2}}]$. The T-matrix does not depend on $k$ because the impurity is localized in real space. This leads to the following LDOS correction
\begin{align}
\delta\rho_{A}(m,\omega) &= \sum_{n=1}^{N} \delta\rho_{A,n}(m,\omega)
= \sum_{n=1}^{N} \cos(2k_{0}m+\tau_{n}(k_{0}))\, \delta\big(\omega+E_{n}(k_{0})\big)\notag \\
\delta\rho_{B}(m,\omega) &= \sum_{n=1}^{N} \delta\rho_{B,n}(m,\omega)= \sum_{n=1}^{N} \cos(2k_{0}m+\tau_{n}(k_{0})+\varphi_{n}(k_{0}))\, \delta\big(\omega+E_{n}(k_{0})\big) \,,
\end{align}
where $\varphi_{n}(k_{0})=\theta_{n}(k_{0})-\theta_{n}(-k_{0})=2\theta_{n}(k_{0})$ under time-reversal symmetry. Besides, we have only considered the conduction bands (i.e., there is no term with $\delta\big(\omega-E_{n}(k_{0})\big)$), and
\begin{align}
V_{n}(\omega)=\frac{V_{n}}{2v_{n}(k_{0})}\frac{1}{\sqrt{(2v_{n}(k_{0}))^{2}+V_{n}^{2}}} \,.
\end{align}
The energy-resolved interference pattern in momentum space behaves similarly to
\begin{align}
\delta\rho_{A}(q,\omega) &= \sum_{n=1}^{N} \delta\rho_{A,n}(q,\omega)
= \sum_{p=-\infty}^{+\infty} \sum_{n=1}^{N} V_{n}(\omega) \frac{\cos(\tau_{k_{0}})-\cos(2k_{0}-q-\tau_{k_{0}}+p2\pi)}{1-\cos(2k_{0}-q+p2\pi)}\, \delta\big(\omega+E_{n}(k_{0})\big)
\notag \\
\delta\rho_{B}(q,\omega) &= \sum_{n=1}^{N} \delta\rho_{B,n}(q,\omega)
= \sum_{p=-\infty}^{+\infty} \sum_{n=1}^{N} V_{n}(\omega) \frac{\cos(\tau_{k_{0}})-\cos(2k_{0}-q-\tau_{k_{0}}+p2\pi)}{1-\cos(2k_{0}-q+p2\pi)}\, e^{-i\varphi_{n}(k_{0})}\, \delta\big(\omega+E_{n}(k_{0})\big)
\,.
\end{align}

When moving along the $q$-periodic paths $\mathscr{C}_{n}$ defined by the constraint $q=2k_{0}$ within each energy band $E_{n}$, we finally access:
\begin{align}
\sum_{n=1}^{N}\oint_{\mathscr{C}_{n}} \frac{dq}{2i\pi}\, \partial_{q} \ln \left[\frac{\delta\rho_{A,n}(q,\omega)}{\delta\rho_{B,n}(q,\omega)}\right]
= \sum_{n=1}^{N} \oint_{\mathscr{C}_{n}} \frac{dq}{2\pi}\, \partial_{q}\, \varphi_{n}(q/2)
= 2 \sum_{n=1}^{N} \oint_{\mathscr{C}_{n}} \frac{dq}{2\pi}\, \partial_{q}\, \theta_{n}(q)
= 2 \sum_{n=1}^{N} W_{n} = 2 W \,.
\end{align}

\end{document}